\documentclass[10pt,journal]{IEEEtran}

\usepackage[hidelinks]{hyperref}
\usepackage{amsmath,amssymb,amsfonts}
\usepackage{mathtools}

\usepackage{graphicx}
\usepackage{xpatch}

\usepackage[caption=false, font=footnotesize]{subfig}

\usepackage[utf8]{inputenc}

\usepackage{sansmath}
\usepackage[defaultmathsizes, italic]{mathastext}

\usepackage{ifdraft}
\usepackage{etoolbox}
\usepackage{xspace}

\usepackage[per-mode=symbol,group-separator={,},detect-all, list-final-separator={, and }]{siunitx}

\usepackage{tikz}
\usetikzlibrary{positioning}
\usetikzlibrary{backgrounds,scopes}
\usetikzlibrary{fit}
\usetikzlibrary{shapes}
\usetikzlibrary{matrix}
\usetikzlibrary{math}
\usetikzlibrary{calc}
\usetikzlibrary{arrows.meta}
\usetikzlibrary{decorations.pathreplacing}
\usetikzlibrary{calligraphy}
\usetikzlibrary{external}
\usetikzlibrary{patterns}

\usepackage{tikzpagenodes}
\usepackage[siunitx]{circuitikz}
\usepackage{pgfplots}
\pgfplotsset{compat=1.18} 
\usepackage{circledsteps}
\newcommand\myCircled[2][]{\ifmmode%
\Circled[fill color=black,inner color=white,#1]{\footnotesize\mathsf{#2}}%
\else%
\Circled[fill color=black,inner color=white,#1]{\footnotesize\sffamily#2}%
\fi%
}

\tikzset{
    font={\fontsize{8pt}{9}\selectfont},
    arrow/.style={-latex}
    }

\usepackage[noabbrev,capitalise]{cleveref}
\Crefname{subsection}{Section}{Sections}
\Crefname{subsubsection}{Section}{Sections}
\Crefname{paragraph}{Section}{Sections}
\Crefname{figure}{Fig.}{Fig.}
\Crefname{table}{Tab.}{Tab.}

\usepackage{microtype}
\usepackage[normalem]{ulem}
\usepackage{tabularx}
\usepackage{booktabs}
\usepackage{multirow}
\usepackage{makecell}
\usepackage{textcomp}
\usepackage[bibstyle=ieee, citestyle=numeric-comp, maxbibnames=15, minbibnames=15, maxcitenames=1, mincitenames=1, doi=false, isbn=false, url=false, 
backend=biber]{biblatex}
\DeclareFieldFormat{journaltitle}{#1\isdot}
\DeclareFieldFormat[book]{title}{{#1\isdot}}

\AtEveryBibitem{
 \clearname{editor}
 \clearfield{month}
  \clearfield{extra}
}
\DeclareSourcemap{
  \maps[datatype=bibtex]{
    \map{
      \step[fieldset=note, null]
    }
  }
}
\DefineBibliographyStrings{english}{%
    andothers = {et al\adddot}
}

\def\nobreakbefore{%
  \relax\ifvmode\else
    \ifhmode
      \ifdim\lastskip > 0pt\relax
        \unskip\nobreakspace
      \fi
    \fi
  \fi
}
\let\oldcite\cite
\renewcommand\cite{\nobreakbefore\oldcite}

\usepackage{enumitem}
\usepackage{balance}

\usepackage{algorithm} 
\usepackage{algpseudocode}

\ifoptionfinal{%
	\usepackage[author={Paul Genssler},color=yellow, final]{pdfcomment}}{%
	\usepackage[author={Paul Genssler},color=yellow, draft]{pdfcomment}}

\makeatletter
\g@addto@macro\@floatboxreset{\centering}
\makeatother

\definecolor{dgreen}{RGB}{0, 139, 0}

\title{EM-Aware Physical Synthesis: Neural Inductor Modeling and Intelligent Placement \& Routing \\ for RF Circuits}

\begin{document}

\author{
Yilun Huang\textsuperscript{\dag}, Asal Mehradfar\textsuperscript{*},  Salman Avestimehr\textsuperscript{*}, Hamidreza Aghasi\textsuperscript{\dag} 

\textsuperscript{*}Electrical Engineering, University of Southern California, Los Angeles, CA, USA \\
\textsuperscript{\dag}Electrical Engineering and Computer Science, University of California Irvine, Irvine, CA, USA \\


\thanks{}

\vspace{-7mm}
\thanks
}



\maketitle
\thispagestyle{empty}
\pagestyle{empty}
\begin{abstract}
This paper presents an ML-driven framework for automated RF physical synthesis that transforms circuit netlists into manufacturable GDSII layouts. While recent ML approaches demonstrate success in topology selection and parameter optimization, they fail to produce manufacturable layouts due to oversimplified component models and lack of routing capabilities. Our framework addresses these limitations through three key innovations: (1) a neural network framework trained on 18,210 inductor geometries with 
frequency sweeps from 1--100~GHz, generating 7.5\,million training samples that predicts inductor Q-factor with $<$2\% error and enables fast gradient-based layout optimization with a 93.77\% success rate in producing high-Q layouts (2) an intelligent P-Cell optimizer that reduces layout area while maintaining design-rule-check (DRC) compliance, and (3) a complete placement and routing engine with frequency-dependent EM spacing rules and DRC-aware synthesis. The neural inductor model demonstrates superior accuracy across 1--100~GHz, enabling EM-accurate component synthesis with real-time inference. The framework successfully generates DRC-aware GDSII layouts for RF circuits, representing a significant step toward automated RF physical design.
\end{abstract}


\section{Introduction}

The design of analog and radio frequency (RF) circuits remains a critical bottleneck in integrated circuit development, with design cycles often spanning months and requiring extensive expert iteration. The complexity arises from continuous parameter spaces, tight coupling between circuit performance and physical layout, and frequency-dependent electromagnetic effects that dominate behavior at millimeter-wave frequencies. Designers must manually adjust component values, run electromagnetic (EM) simulations, and iterate layout to meet specifications---a process that involves repeated cycles of schematic optimization, layout implementation, and parasitic extraction, creating a fundamental scalability challenge for communications, radar~\cite{xuyang}, and sensing applications.

Recent advances in machine learning, including evolutionary and 
surrogate-based methods~\cite{liu2011synthesis, passos2023pacosyt, liu2014gaspad, aicircuitjournal}, 
have demonstrated promise for automating circuit design at the schematic level. Frameworks like FALCON~\cite{mehradfar2025falcon} achieve high accuracy in topology selection and parameter optimization through neural networks and differentiable optimization~\cite{karahan2024deeplearning, zhang2024intelligent, xiao2021inverse}. Other approaches, including genetic programming, diffusion models, and 
LLM-based methods~\cite{mcconaghy2011trustworthy, azevedo2025diffusion, liu2025llmbased}, explore reinforcement learning for circuit sizing~\cite{DATE_AutoCkt,gcnrl} and Bayesian optimization for performance prediction~\cite{bayes}. These methods successfully address the schematic-level optimization problem, generating component values that meet electrical specifications. However, they face a critical limitation: the inability to generate physical layouts required for fabrication. 
\begin{figure}[t]
    \centering    \includegraphics[width=0.38\textwidth]{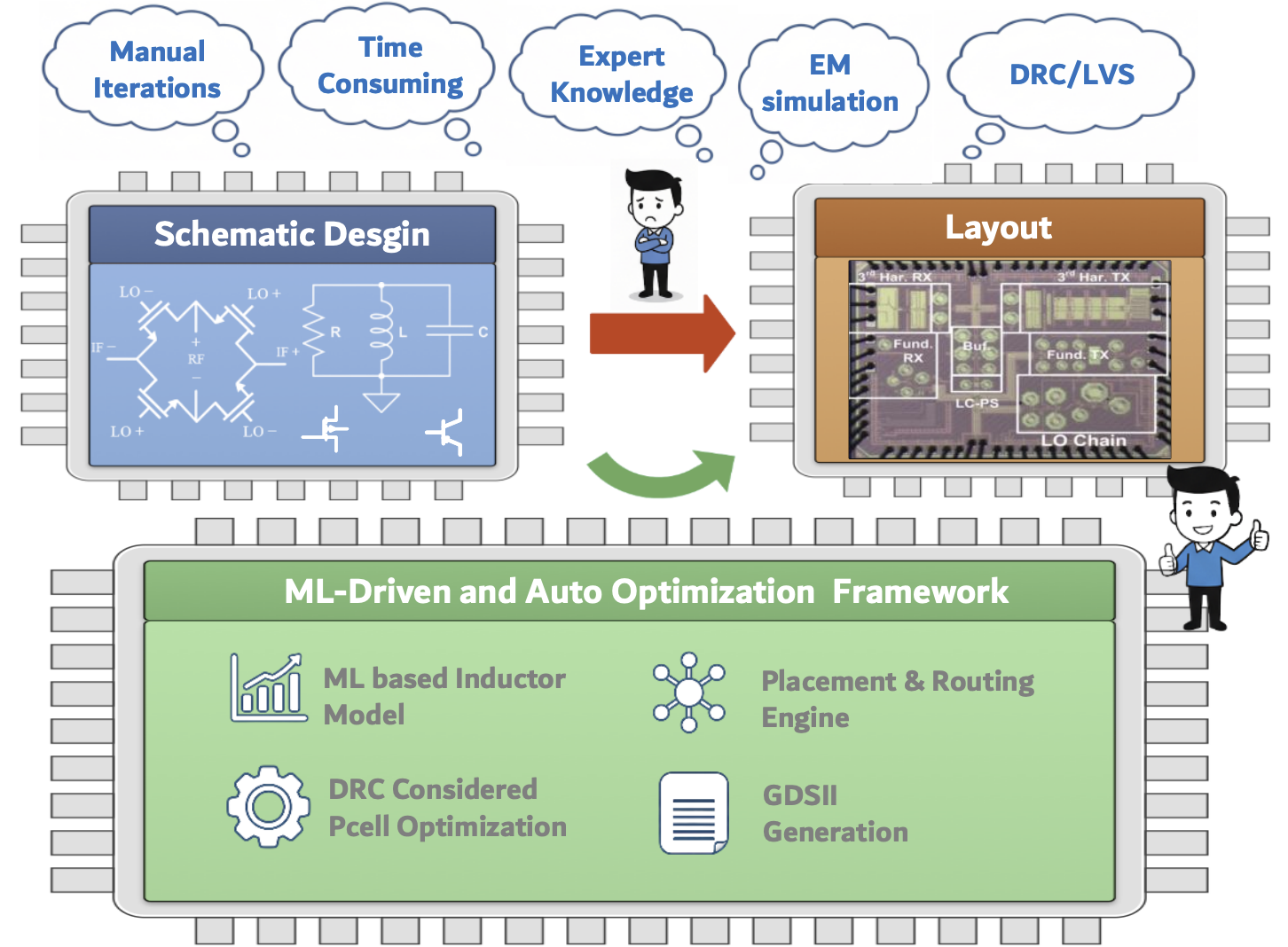}
    \caption{Proposed RF synthesis framework for automated netlist-to-GDSII.}
    \vspace{-5mm}
    \label{fig:intro}
\end{figure}
The physical implementation gap poses severe practical challenges. While FALCON effectively introduces layout awareness through differentiable cost functions and achieves good accuracy at RF and low-mm-wave frequencies, its analytical passive models become less reliable at higher mm-wave bands where full electromagnetic effects dominate. Traditional analytical inductor models such as Wheeler's approximation \cite{wheeler} and modified current sheet methods \cite{sheet} produce Q-factor errors at millimeter-wave frequencies, limiting the accuracy of layout-aware optimization when using these simplified approaches. Furthermore, existing ML frameworks output only optimized component parameters without generating the placement, routing, or GDSII files \cite{mirhoseini2021graph, chen2021magical}, forcing designers to manually translate these into layouts.

This work bridges ML-based circuit optimization and manufacturable physical layouts through a complete RF synthesis framework in Fig.~\ref{fig:intro}. By replacing inaccurate analytical models with data-driven EM predictions and automating the entire layout generation process, we enable direct netlist-to-GDSII synthesis. The framework makes three key contributions: (1)~EM-accurate ML inductor modeling, (2)~EM-aware placement and routing and (3)~complete netlist-to-GDSII. The framework is validated on 22-nm CMOS RF circuits operating from 1--100~GHz, demonstrating the first complete path from circuit netlist to layout using ML-driven component modeling.
\section{System Overview}

\begin{figure}[t]
    \centering
    \includegraphics[width=0.4\textwidth]{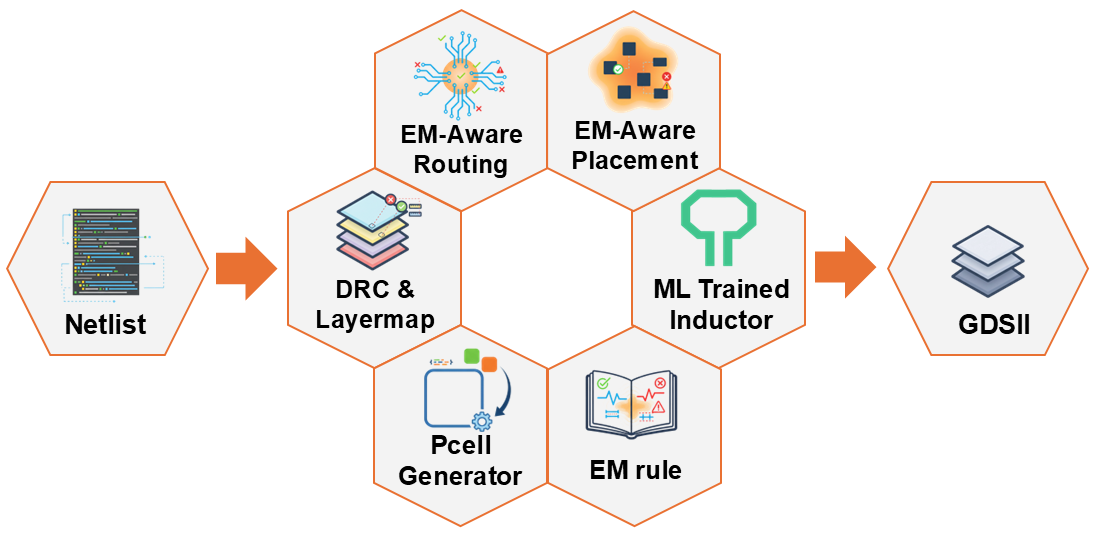}
    \vspace{-0.15in}
    \caption{The proposed pipeline of GDS from netlist generation which captures various types of constraints without an specific order. }
    \vspace{-1mm}
    \label{fig:sys}
\end{figure}

The proposed framework transforms circuit netlists into manufacturable GDSII layouts through an integrated synthesis pipeline, as illustrated in Fig.~\ref{fig:sys}. Beginning with a circuit netlist specification, the framework generates complete physical implementations by orchestrating multiple specialized synthesis engines with electromagnetic awareness.

The \textbf{ML-trained Inductor} model addresses the fundamental limitation of analytical inductor equations by leveraging data-driven predictions derived from full-wave electromagnetic simulations. This approach enables accurate inductor synthesis across the 1--100~GHz frequency range, capturing frequency-dependent substrate losses, skin effect, and parasitic coupling that analytical models fail to represent.

Component instantiation for capacitors and resistors is handled by the \textbf{PCell Generator}, which systematically explores metal stack configurations to minimize layout area while satisfying electrical specifications and maintaining DRC.

The \textbf{EM-aware Placement} engine positions components according to spacing constraints encoded in the \textbf{EM Rule} database. These rules implement frequency-dependent minimum separation requirements that mitigate performance degradation arising from substrate coupling and electromagnetic interference—effects that become critical at mm-wave.

Interconnect synthesis is performed by the \textbf{EM-aware Routing} engine, which consults both the foundry-specific \textbf{Technology Layermap} for appropriate layer assignments and \textbf{DRC} constraints for design rule compliance. The routing algorithm incorporates spacing margins to reduce violations that would otherwise necessitate manual layout iteration.

The final \textbf{Generator} module assembles component geometries and routing topologies into hierarchical GDSII files.

\begin{figure}[!ht]
    \centering
    \includegraphics[width=0.38\textwidth]{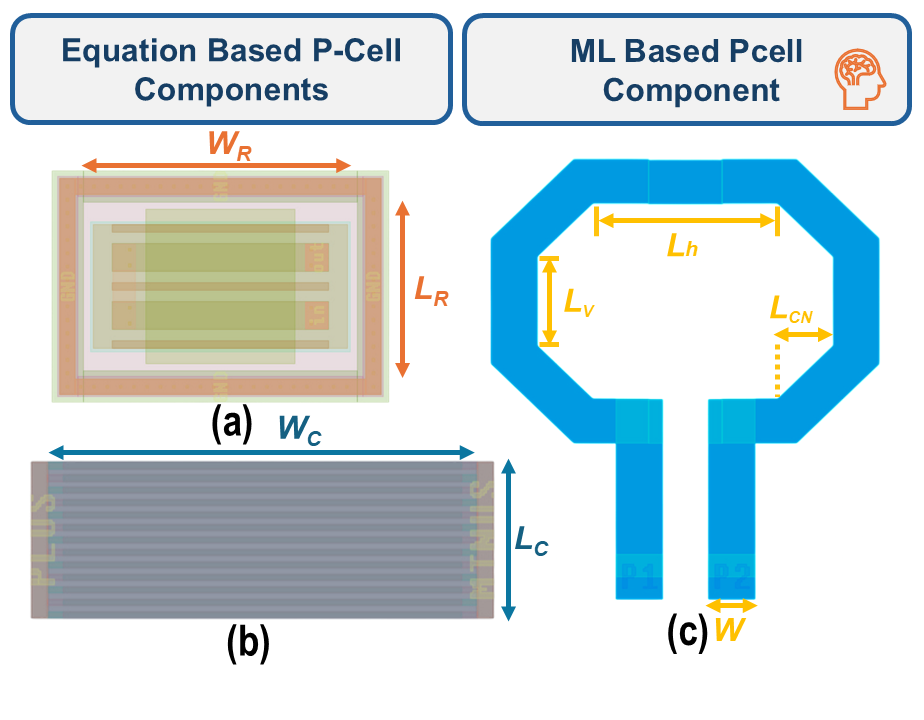}
    \vspace{-0.15in}
    \caption{(a) N+ Diffusion Resistor (b) Analog Passive Metal-Oxide-Metal Capacitor (c) ML Based Inductor }
    \label{fig:inductor}
\end{figure}
\section{ML-Based Q-Factor Prediction and Layout Optimization}

\subsection{Data Preprocessing}
The dataset used in this study comprises approximately 7.5\,million samples, each annotated with a set of geometric and electrical parameters along with the corresponding quality factor ($Q$). The available parameters include the target specifications $(f, W, L)$ and layout-related variables $(L_v, L_h, L_{CN})$, where $W$, $L_v$, $L_h$, and $L_{CN}$ denote the inductor width and layout dimensions (all expressed in $\mu$m), as illustrated in Fig.~\ref{fig:inductor}. $L$ is the inductance in pH, and $f$ is the target frequency in GHz.

\textbf{Feature Selection and Cleaning.}  
Raw simulation files were parsed to extract $f$, $W$, $L$, $L_v$, $L_h$, $L_{CN}$, and $Q$. Entries with non-physical values were removed, and the cleaned data were aggregated into a unified dataset.

\begin{figure}[!h]
    \centering
\includegraphics[width=0.8\columnwidth]{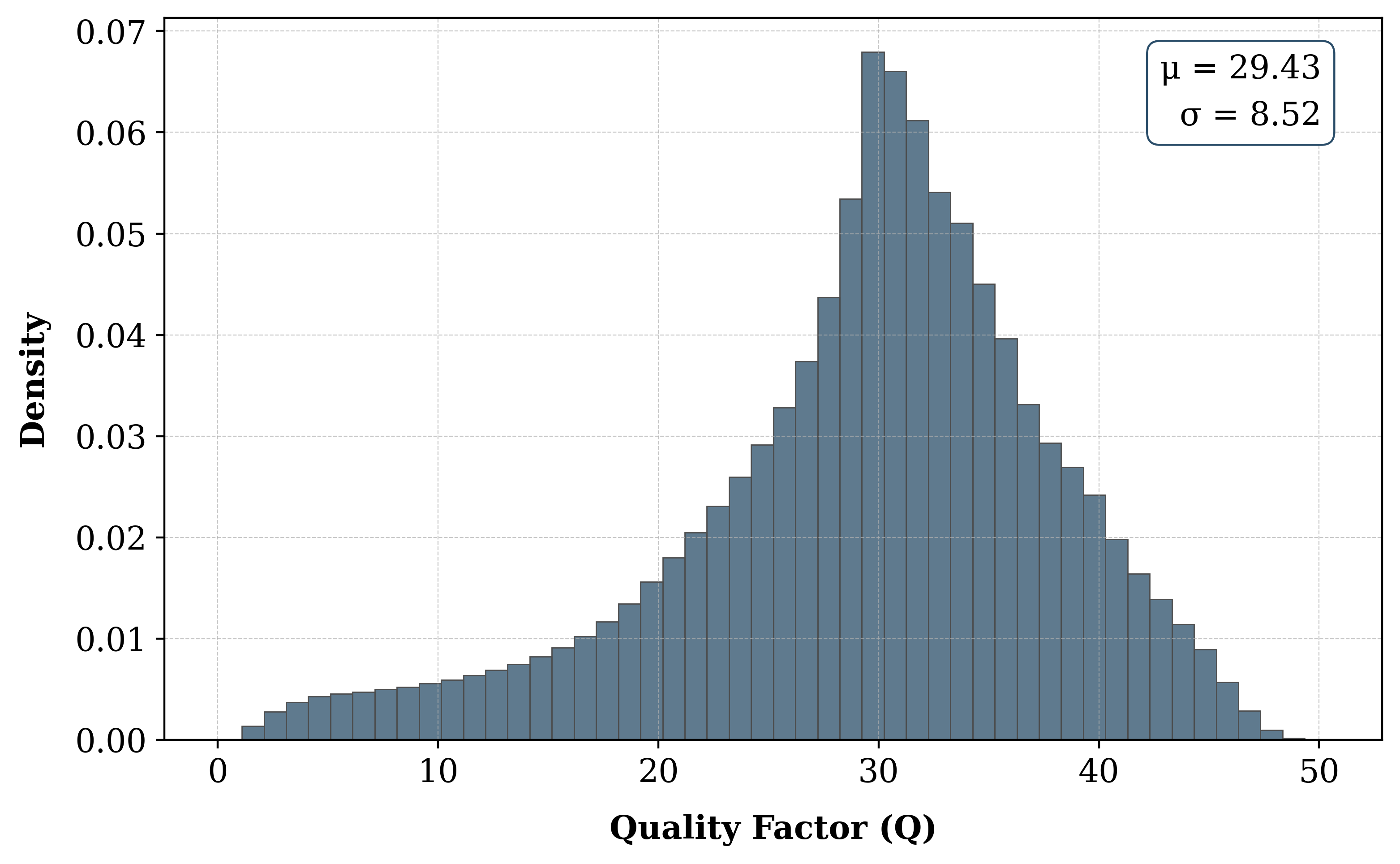}
    \vspace{-0.1in}
    \caption{Distribution of the inductor quality factor $Q$ across the dataset.}
    \label{fig:q_distribution}
\end{figure}

\textbf{Distribution Visualization.}  
Fig.~\ref{fig:q_distribution} shows the $Q$ distribution, spanning 0–50 with annotated mean and standard deviation.

\textbf{Normalization.}  
Design and layout variables $(f, W, L, L_v, L_h, L_{CN})$ were normalized using training-set statistics as $x_i' = (x_i - \mu_i)/\sigma_i$, where $\mu_i$ and $\sigma_i$ are the feature mean and standard deviation. The quality factor $Q$ was left unnormalized to retain physical interpretability.

\textbf{Data Splitting.}  
The dataset was divided into training, validation, and test sets (80:10:10) using a fixed seed. These subsets were used for learning, hyperparameter tuning, and final evaluation, respectively.

\subsection{Forward Model Training and Evaluation}

We formulate quality-factor prediction as a supervised regression task. Given inputs $\mathbf{x} = (f, W, L, L_v, L_h, L_{CN})$, the goal is to learn $Q_\theta(\mathbf{x})$, parameterized by $\theta$, that predicts the corresponding quality factor $Q$. The model minimizes the mean squared error (MSE), $\mathcal{L}_\text{MSE} = \frac{1}{N}\sum_{i=1}^{N}(Q_\theta(\mathbf{x}_i) - Q_i)^2$, where $N$ is the number of samples and $Q_i$ is the ground-truth label. An MLP model $Q_\theta$ is designed with ten hidden layers of widths $\{256, 256, 256, 128, 128, 128, 64, 64, 64, 32\}$, each followed by ReLU and LayerNorm for stability. A final linear layer with Softplus activation, $\text{Softplus}(z) = \log(1 + e^z)$, ensures non-negative outputs consistency with physical constraints.

Training uses Adam (learning rate $0.001$, batch size 16,384) with \texttt{ReduceLROnPlateau} to lower the rate when validation loss plateaus. The model trains for up to 300~epochs with early stopping, and the checkpoint with the lowest validation MSE is used for final testing.

\begin{table}[h]
\centering
\caption{Evaluation of forward $Q$ prediction on held-out test set.}
\label{tab:regression_metrics}
\begin{tabular}{lc}
\toprule
\textbf{Metric} & \textbf{Value} \\
\midrule
Mean Absolute Error (MAE)       & 0.419 \\
Mean Squared Error (MSE)        & 0.423 \\
Root Mean Squared Error (RMSE)  & 0.65 \\
R-squared ($R^2$ Score)         & 0.994 \\
Mean Absolute Percentage Error (MAPE) & 1.36\% \\
\bottomrule
\end{tabular}
\end{table}
The trained model achieves the results in Table~\ref{tab:regression_metrics} on the held-out test set, reflecting the prediction error between the model’s forward estimates and the actual simulated $Q$ values without any layout optimization. This strong agreement confirms that the model captures the relationship between design parameters and $Q$, providing a solid foundation for inverse design.

\vspace{-0.15in}
\subsection{Inverse Layout Design via Gradient Reasoning}
We adopt a differentiable inverse design strategy inspired by the gradient reasoning method proposed in FALCON~\cite{mehradfar2025falcon} to maximize the quality factor $Q$ by tuning layout parameters. Given fixed target specifications—operating frequency $f$, width $W$, and inductance $L$—our goal is to optimize the layout variables: vertical length $L_v$, horizontal length $L_h$, and center length $L_{CN}$. The inverse design task is formulated as $(L_v^*, L_h^*, L_{CN}^*) = \arg\max_{L_v, L_h, L_{CN}} Q_\theta(f, W, L, L_v, L_h, L_{CN})$, where $Q_\theta$ denotes the frozen prediction model trained in the forward regression task. The optimization is subject to physical box constraints: 
$W + 2 \leq L_v \leq 100~(\mu\text{m})$, 
$2W + 4 \leq L_h \leq 100~(\mu\text{m})$, and 
$1 \leq L_{CN} \leq 50~(\mu\text{m})$.

We initialize each optimization run from a feasible design starting point of $(L_v, L_h, L_{CN}) = (40, 40, 20)\,\mu\text{m}$. At every step, the candidate values are concatenated with fixed $(f, W, L)$ and normalized using the same training-set statistics used during forward modeling. The normalized vector is passed to the frozen MLP to produce the predicted $Q$, and the loss function is defined as the negative prediction, $\mathcal{L} = -Q_\theta(f, W, L, L_v, L_h, L_{CN})$.

We use the Adam optimizer with a learning rate of 0.01 to iteratively update $(L_v, L_h, L_{CN})$, treating them as differentiable parameters. To enforce layout validity and manufacturability, we clamp these variables to their valid physical ranges after each update. By default, we perform optimization for 3000 steps; however, the stopping criterion is {user-configurable}—for example, optimization may be terminated early if a target threshold for $Q$ is reached. {Importantly, these optimization settings can be modified at inference time without retraining the model}, allowing flexible adaptation to diverse design objectives.

To reduce EM simulation cost, inverse design performance is evaluated on 1000 randomly selected samples from the held-out test set. Rather than relying solely on regression metrics, we assess the \emph{success rate}, defined as the percentage of cases where the simulated quality factor $Q$ exceeds a practical threshold. Using $Q$$>$10 as the design goal, the method achieves a success rate of {93.77\%}, indicating that the learned optimization strategy consistently generates high-performing, physically valid layouts. {Each optimization instance runs in under one second on a standard MacBook CPU}, showcasing the method's computational efficiency and real-world deployability. The source code is available at: \url{https://github.com/AsalMehradfar/InductorModeling}
\section{Layout Generation}
\label{sec:layout}

\subsection{Intelligent PCell Optimization}

Foundry PCells prioritize flexibility over area. We perform automated search to minimize area while meeting electrical targets and DRC margins.

\textbf{MOM Capacitor.} We use density model $C_{\mathrm{pF}}=\rho_{\text{stack}}\,A_{\mathrm{eff}}\times10^{-3}$, where $\rho_{\text{stack}}$ is the capacitance density (fF/$\mu\mathrm{m}^2$) for a specific metal stack configuration and $A_{\mathrm{eff}}=W\cdot L$. For 22\,nm CMOS at 1.8\,V,Stacks span $\ge3$ layers. Given $C_{\text{target}}$, the optimizer enumerates stack choices and $(W,L)$ pairs ($1.0\le L\le 60.0~\mu\mathrm{m}$, $1.0\le W\le 330.0~\mu\mathrm{m}$), selecting minimum-area solution within $\pm0.5\%$ of target.

\textbf{Resistor.} For poly resistors we model $R_{\text{stripe}} = R_s(L/W) + 2R_{\text{end}}/W$ with $R_s$ (sheet resistance) and $R_{\text{end}}$ (contact resistance) based on PDK values. Series/parallel tiling yields $R_{\text{total}}=(N_s/N_p)\,R_{\text{stripe}}$. We search $1\le N_s,N_p\le 64$ under $0.36\le W\le 3.72~\mu\mathrm{m}$ and $0.40\le L\le 50.0~\mu\mathrm{m}$, choosing smallest-area design meeting target.
\vspace{-0.1in}
\subsection{Multi-Objective Placement Optimization}

We optimize device positions $\mathbf{p}=\{(x_i,y_i,\theta_i)\}_{i=1}^N$ where $(x_i,y_i)$ is the coordinate and $\theta_i\in\{0^\circ,90^\circ,180^\circ,270^\circ\}$ is rotation:
\begin{equation}
\min_{\mathbf{p}} \quad C(\mathbf{p}) = \text{HPWL}(\mathbf{p}) + K \sum_{1 \le i < j \le N} A_{\text{overlap}}(B_i,B_j),
\label{eq:placement}
\end{equation}
where HPWL (half-perimeter wirelength) measures interconnect length, $B_i$ are device bounding boxes, and $K=10^4$ penalizes overlaps.

\begin{algorithm}[h]
\caption{Multi-Objective Device Placement}
\label{alg:placement}
\footnotesize
\setlength{\belowcaptionskip}{-5pt}
\begin{algorithmic}[1]
\Require Devices $\mathcal{D}$, netlist $\mathcal{G}$, spacing $S_{\min}(f)$, guard $\Delta$
\Ensure Placement $\mathbf{p}$
\State Order by connectivity $d_i=|\{n: i\in P_n\}|$; initial row placement
\For{$t=1$ to $T_{\max}$} \Comment{Local search}
    \State Move: swap/translate; $C(\mathbf{p})=\text{HPWL}+K\sum A_{\text{overlap}}$
    \State Accept if $C_{\text{new}}<C_{\text{curr}}$, else revert
\EndFor
\For{each device $i$} \Comment{Rotation}
    \State Score $\theta\in\{0^\circ,90^\circ,180^\circ,270^\circ\}$: $S_\theta=\min_p\max_d(L_{\text{free}}-L_{\text{need}})$
    \State Select $\theta_i^*=\arg\max S_\theta$
\EndFor
\end{algorithmic}
\end{algorithm}

\begin{figure*}[t!]
    \centering
    \includegraphics[width=0.9\textwidth]{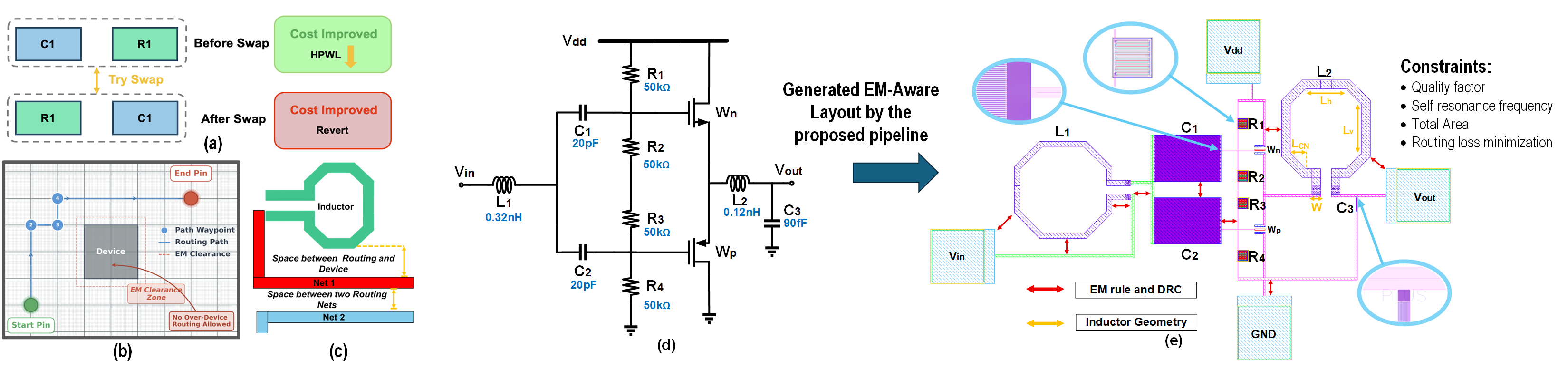}
    \vspace{-0.15in}
    \caption{(a) Adjacent-swap local move. Swap accepted only if augmented objective (criticality-weighted HPWL plus soft separation) improves. (b) Grid-based A* routing on 0.1\,$\mu$m grid avoiding device obstacles with Manhattan routing. (c) Router spacing policy. Top: routing-to-device isolation (no over-cell). Bottom: inter-net spacing. Both obey frequency-aware targets. (d) Designed Class-B PA schematic (e) Generated layout of Class-B PA by the proposed pipeline.}
    \vspace{-3mm}
    \label{fig:routing}
\end{figure*}

Devices are ordered by connectivity degree $d_i$. We perform adjacent swaps (exchanging neighboring devices) and small translations, as shown in Fig.~\ref{fig:routing}(a). A move is accepted only if cost improves, considering criticality-weighted HPWL (high-priority nets weighted by $w_n\!\ge\!1$) and soft penalties for tight spacing. All spacings use 10--20\% DRC guard-band $\Delta$ (heuristically $\sim$10\,\textmu m at $<5$\,GHz, $\sim$30\,\textmu m at $>40$\,GHz). For rotation, we score by $S_{\theta}=\min_{p\in\text{Pins}}\max_{d\in D_p}(L_{\text{free}}-L_{\text{need}})$, where $L_{\text{free}}(p,d,\theta)$ is available routing space from pin $p$ in direction $d\in\{L,R,U,D\}$ (left/right/up/down) at rotation $\theta$, and $L_{\text{need}}=\text{dist}(p,\partial B)+m_{\text{margin}}$ is required escape distance from device boundary $\partial B$ plus margin. NMOS pins (Gate/Drain/Source) are recognized, preferring vertical S/D escapes.


\vspace{-0.1in}
\subsection{Constraint-Driven Global Routing}

Routing uses A* on grid $\mathcal{G}=\{(x,y,\ell): x,y\in 0.1~\mu\text{m}\cdot\mathbb{Z},\; \ell\in\{\text{M1,QA,QB}\}\}$ where $(x,y)$ are coordinates with 0.1\,$\mu$m resolution and $\ell$ is the metal layer, with cost function:
\begin{equation}
f(p)=g(p)+h(p)+\lambda_{\text{layer}}\cdot\mathbb{1}_{[\text{layer switch}]},
\end{equation}
where $g(p)$ is path cost from start, $h(p)=\|p-t\|_1$ is Manhattan distance to target $t$, and $\lambda_{\text{layer}}=10\cdot g$ discourages vias. Net topology is minimum spanning tree (MST) minimizing $\sum_{e\in E_n}\|e\|_1$. Fig.~\ref{fig:routing}(b) illustrates grid-based routing with obstacle avoidance.
\begin{algorithm}[h]
\caption{EM-Aware A* Global Routing}
\label{alg:routing}
\footnotesize
\setlength{\belowcaptionskip}{-5pt}
\begin{algorithmic}[1]
\Require Pins $\mathcal{P}_n$, grid $\mathcal{G}$, layers $\mathcal{L}$, devices $\{B_i\}$, DRC
\Ensure Routes $\mathcal{R}_n$
\State $T_n \gets \mathrm{MST}(\mathcal{P}_n)$; \textbf{for each} edge $(s,t)\!\in\!T_n$:
\State \hspace{0.5em}$Q\!\gets\!\{(f(s),s)\}$ (min-heap on $f$), $C\!\gets\!\emptyset$, $g(s)\!\gets\!0$; step $h\!=\!0.1\,\mu$m
\While{$Q\neq\emptyset$}
  \State pop best $p$ from $Q$; \textbf{if} $\|p-t\|_1<h$ \textbf{then} emit path $\to$ GDSII; \textbf{continue}
  \State add $p$ to $C$
  \For{$d\in\{L,R,U,D\}$} \Comment{in-plane moves}
    \State $p'\!\gets\!p+h\cdot d$; \textbf{if} $p'\in C$ \textbf{or} DRC-violate$(p')$ \textbf{then continue}
    \State $g'\!\gets\!g(p)+h$; $f'\!\gets\!g'+\|p'-t\|_1$; relax/insert $(f',p')$ into $Q$
  \EndFor
  \For{$\ell'\in\mathcal{L}\setminus\{\ell\}$} \Comment{via moves}
    \State $p_v\!\gets\!(x,y,\ell')$; $g'\!\gets\!g(p)+10\,g(p)$; $f'\!\gets\!g'+\|p_v-t\|_1$; relax/insert $(f',p_v)$ into $Q$
  \EndFor
\EndWhile
\end{algorithmic}
\end{algorithm}

\textbf{No-over-cell.} Expanded segments $R_{\text{seg}}\oplus\mathcal{B}(s_{\text{same}}/2)$ must satisfy $R_{\text{seg}}^{\text{exp}} \cap B_i = \emptyset, \forall i$, except pin corridors along $d\in\{L,R,U,D\}$.

\textbf{EM-aware spacing.} Devices and routes are dilated by required clearance (Fig.~\ref{fig:routing}(c)); grid cells inside halos are blocked. Clearances distinguish routing-to-device (no-over-cell) vs.\ inter-net spacing (crosstalk prevention) and scale with frequency. All rules use 10--20\% guard-bands.

\textbf{Pin escape.} Two-phase: (1) straight escape selecting $d^*=\arg\max_d L_{\text{free}}(p,d)$ with maximum free run, or (2) dogleg (orthogonal-then-primary) if blocked. Both validate no-over-cell and spacing constraints.
\vspace{-0.15in}
\subsection{GDSII Generation}

The framework generates hierarchical GDSII files through Python-based PCell assembly using the \texttt{gdstk} library, interfacing with foundry PDKs. Notably, in the Class-B PA design example (Fig.~\ref{fig:routing}(d, e)), transistors are represented as simplified three-pin boxes instead of transistor layouts. This simplification represents a current limitation, as the framework does not yet fully support the complex mirroring and symmetry requirements inherent to RF/analog transistor layout design.

\begin{table}[t]
\centering
\caption{Comparison of This Work with Prior Works: Focus on Layout and EM Automation Capabilities}
\label{tab:related_work_comparison}
\footnotesize
\begin{tabular}{lccccc}
\toprule
\textbf{Method} & 
\makecell{\textbf{Layout}\\\textbf{Gen.}} & 
\makecell{\textbf{EM-}\\\textbf{Aware}} & 
\makecell{\textbf{RF/}\\\textbf{mm-W}} & 
\makecell{\textbf{Foundry}\\\textbf{PDK}} & 
\makecell{\textbf{Public}\\\textbf{D/C}} \\
\midrule
CktGNN~\cite{dong2023cktgnn} & $\times$ & $\times$ & $\times$ & $\times$ & \checkmark/\checkmark \\
LaMAGIC~\cite{chang2024lamagic} & $\times$ & $\times$ & $\times$ & $\times$ & $\times$/$\times$ \\
AnalogCoder~\cite{lai2025analogcoder} & $\times$ & $\times$ & $\times$ & $\times$ & \checkmark/\checkmark \\
GCN-RL~\cite{gcnrl} & $\times$ & $\times$ & $\times$ & \checkmark & $\times$/$\times$ \\
Cao et al.~\cite{cao2022domain} & $\times$ & $\times$ & $\times$ & \checkmark & $\times$/$\times$ \\
BO-SPGP~\cite{bayes} & $\times$ & $\times$ & $\times$ & \checkmark & $\times$/$\times$ \\
AICircuit~\cite{Mehradfar2024AICircuit} & $\times$ & $\times$ & \checkmark & \checkmark & \checkmark/\checkmark \\
Krylov et al.~\cite{ICML_Analog} & $\times$ & $\times$ & $\times$ & $\times$ & \checkmark/\checkmark \\
LayoutCopilot~\cite{liu2025layoutcopilot} & \checkmark & $\times$ & $\times$ & \checkmark & $\times$/$\times$ \\
AutoCkt~\cite{DATE_AutoCkt} & $\times$ & $\times$ & $\times$ & \checkmark & $\times$/$\times$ \\
CAN-RL~\cite{Li2021attention} & \checkmark & $\times$ & $\times$ & \checkmark & $\times$/$\times$ \\
\midrule
\textbf{This Work} & \textbf{\checkmark} & \textbf{\checkmark} & \textbf{\checkmark} & \textbf{\checkmark} & \textbf{\checkmark/\checkmark} \\
\bottomrule
\end{tabular}
\vspace{1mm}
\begin{flushleft}
\scriptsize
\textit{Layout Gen.}: Automated layout synthesis. \textit{EM-Aware}: EM effects in layout. \textit{RF/mm-W}: RF/mm-wave layout. \textit{Foundry PDK}: Uses foundry PDKs. \textit{Public D/C}: Dataset/Code (format: Data/Code).
\vspace{-6mm}
\end{flushleft}
\end{table}
\section{Conclusion}
This work presents an EM-aware, ML-driven physical-synthesis flow that maps RF netlists directly to GDSII. A neural inductor surrogate trained on HFSS simulations achieves \mbox{$<$2\%} Q-factor error across 1--100\,GHz inference and enables fast gradient-based layout optimization with a 93.77\% success rate of high-Q layouts, enabling EM-accurate passive synthesis. An intelligent PCell optimizer reduces passive area, and a frequency-aware placement-and-routing engine enforces technology rules and spacing to yield hierarchical, DRC-compliant layouts in a 22\,nm CMOS process. Next, we will broaden coverage across circuit blocks and technology nodes, strengthen EM-in-the-loop cost modeling and post-layout verification, and enhance robustness to variation and reliability, while exploring tighter integration with higher-level design automation and adaptive rule learning—pushing toward a scalable, production-ready path from schematic intent to tapeout.

\section*{Acknowledgment}
The authors would like to thank Global Foundries for technology access. This project is partially sponsored under National Science Foundation Grant $\#$2443820.

\setlength{\itemindent}{-\biblabelsep}

\renewcommand*{\bibfont}{\footnotesize}
\renewcommand*{\UrlFont}{\rmfamily}
\setlength{\biblabelsep}{\labelsep}
\setlength{\bibhang}{2pt}
\printbibliography

\end{document}